\magnification=1200
\parskip = 10pt plus 5pt
\baselineskip=16pt
\input amssym.def
\input amssym
\topglue 100pt
\line{\hfil {November 15, 1998}}
\pageno = 0
\footline={\ifnum \pageno < 1 \else \hss \folio \hss \fi}
\vskip 60pt
\centerline{\bf The Genetic Code as a Chronicle of Early Events in the
Origin of Life}
\vskip 0.7in
\centerline{\bf Brian K. Davis}
\vskip 0.5in
\centerline{Research Foundation of Southern California Inc.}
\centerline{5580 La Jolla Boulevard, La Jolla CA 92037, U.S.A.}
\vskip 0.4in
\noindent
Molecular evidence regarding the genetic code has been examined and the
findings on the nature of the early events responsible for the amino acid
distribution pattern in the code are reported.

\vfill\eject

\noindent
Molecular evidence on the genetic code has been examined with the aim of
identifying early events responsible for the amino acid distribution pattern
in the code. A tree of codon assignments resulted that was rooted in the
four N-fixing amino acids (Asp. Glu, Asn, Gln) and the sixteen base triplets
of the NAN set (A, adenine; N, any base).  This locally phased (commaless)
code appeared to have arisen from ambiguous translation of a poly(A)
collector strand in a surface reaction network. Copolymerisation of these
amino acids produced polyanionic oligopeptides able to anchor uncharged amide 
residues to a positively charged mineral surface. The first genes were required
to increase efficiency of N-fixation relying on oligopeptides. They were
probably template strands spliced into tRNA. Expansion of the code was
conditional on initiation at the $5^\prime$-base of a translated sequence.
It reduced the risk of mutation to an unreadable codon and led to 
incorporation of increasingly hydrophobic residues. As codons of the
NUN set (U, uracil) were assigned most slowly, they received the most
non-polar amino acids. The folded proteins ferredoxin and Gln synthetase
evidently originated in early- and mid-expansion phase, respectively.
Surface metabolism ceased by the end of code expansion, as cells bounded by
a proteo-phospholipid membrane had emerged. Incorporation of positively 
charged and aromatic amino acids followed. They entered the code by
codon capture. Synthesis of efficient enzymes with acid-base catalysis was 
then possible. Formation of class I and II aminoacyl-tRNA synthetases
was attributed to this stage. The first cells formed in approximately
7 million yr, an interval less than the ocean circulation cycle for
passage through hot submarine vents. It was estimated the Standard Code evolved
within 20 million yr. These studies on the genetic code provide empirical
evidence that a surface reaction network, centred on C-fixing autolcatalytic
cycles, rapidly led to a cellular life on Earth.

\end